\documentclass{emulateapj}
\usepackage{amsmath}
\usepackage{txfonts}
\usepackage{natbib}
\usepackage{graphicx}
\usepackage{color}
\usepackage{multirow}
\usepackage{array}
\usepackage{hyperref}

\def\ga{\,\,\raise0.14em\hbox{$>$}\kern-0.76em\lower0.28em\hbox
{$\sim$}\,\,}
\def\la{\,\,\raise0.14em\hbox{$<$}\kern-0.76em\lower0.28em\hbox
{$\sim$}\,\,}

\shorttitle{Merger ejecta as obstacles to GRB jets}

\shortauthors{Just et al.}

\begin{document}

\title{Neutron-star merger ejecta as obstacles to neutrino-powered jets of gamma-ray bursts}

\author{O.~Just\altaffilmark{1,2}, M.~Obergaulinger\altaffilmark{3}, H.-T.~Janka\altaffilmark{1},
  A.~Bauswein\altaffilmark{4,5}, and N.~Schwarz\altaffilmark{1,6}}

\altaffiltext{1}{Max-Planck-Institut f\"ur Astrophysik, Karl-Schwarzschild-Str.~1, 
85748 Garching, Germany}
\altaffiltext{2}{Max-Planck/Princeton Center for Plasma Physics (MPPC)}
\altaffiltext{3}{Departament d{\'{}}Astronomia i Astrof{\'i}sica, Universitat de Val{\`e}ncia, 
  Edifici d{\'{}}Investigaci{\'o} Jeroni Mu{\~n}oz, C/ Dr.~Moliner, 50, E-46100 Burjassot (Val{\`e}ncia), Spain}
\altaffiltext{4}{Department of Physics, Aristotle University of Thessaloniki, 54124 Thessaloniki, Greece}
\altaffiltext{5}{Heidelberger Institut f\"ur Theoretische Studien, Schloss-Wolfsbrunnenweg 35, 69118~Heidelberg, Germany}
\altaffiltext{6}{Physik Department, Technische Universit\"at M\"unchen, James-Franck-Stra\ss e 1, 85748 Garching, Germany}
\email{ojust@mpa-garching.mpg.de, thj@mpa-garching.mpg.de}

\begin{abstract}
  We present the first special relativistic, axisymmetric hydrodynamic simulations of black
  hole-torus systems (approximating general relativistic gravity) as remnants of binary-neutron star
  (NS-NS) and neutron star-black hole (NS-BH) mergers, in which the viscously driven evolution of
  the accretion torus is followed with self-consistent energy-dependent neutrino transport and the
  interaction with the cloud of dynamical ejecta expelled during the NS-NS merging is taken into
  account. The modeled torus masses, BH masses and spins, and the ejecta masses, velocities, and
  spatial distributions are adopted from relativistic merger simulations. We find that energy
  deposition by neutrino annihilation can accelerate outflows with initially high Lorentz factors
  along polar low-density funnels, but only in mergers with extremely low baryon pollution in the
  polar regions. NS-BH mergers, where polar mass ejection during the merging phase is absent,
  provide sufficiently baryon-poor environments to enable neutrino-powered, ultrarelativistic jets
  with terminal Lorentz factors above 100 and considerable dynamical collimation, favoring short
  gamma-ray bursts (sGRBs), although their typical energies and durations might be too small to
  explain the majority of events. In the case of NS-NS mergers, however, neutrino emission of the
  accreting and viscously spreading torus is too short and too weak to yield enough energy for the
  outflows to break out from the surrounding ejecta shell as highly relativistic jets. We conclude
  that neutrino annihilation alone cannot power sGRBs from NS-NS mergers.
\end{abstract}


\section{Introduction}

\setlength{\tabcolsep}{1.6pt}
\begin{table*}
  \centering
    \caption{\label{tab:jetmodels}Model Properties and Results}
    \begin{tabular}{l cccccccccccccccccc}
      \tableline \tableline 
                    &             &                   &                   &                      &                         &                    &                        &                          &                              &                                   &                                   &                           &                           &                            &                         &                     &                          & \vspace{-2.5mm}                    \\
      Name          & $M_1/M_2$   & $M_{\mathrm{BH}}$ & $A_{\mathrm{BH}}$ & $m_{\mathrm{torus}}$ & $\alpha_{\mathrm{vis}}$ & $m_{\mathrm{env}}$ & $m_{\mathrm{env,pol}}$ & $\bar{v}_{\mathrm{env}}$ & $\bar{v}_{\mathrm{env,pol}}$ & $E_{\mathrm{ann}}^{\mathrm{pol}}$ & $E_{\mathrm{ann}}^{\mathrm{iso}}$ & $\bar\eta_{\mathrm{\nu}}$ & $\bar\eta_{\mathrm{ann}}$ & $T_{\mathrm{NDAF}}$ & $T^{90}_{\mathrm{ann}}$ & $E_{\Gamma>10/100}$ & $\theta_{\Gamma>10/100}$ & $E_{\Gamma>10/100}^{\mathrm{iso}}$ \\
                    & $[M_\odot]$ & $[M_\odot]$       &                   & $[M_\odot]$          &                         & $[10^{-3}M_\odot]$ & $[m_{\mathrm{dyn}}]$   & [c]                      & [c]                          & [$10^{49}\,$erg]                  & [$10^{49}\,$erg]                  & [\%]                      & [\%]                      & [ms]                & [ms]                    & [$10^{48}\,$erg]    & [$^\circ$]               & [$10^{50}\,$erg]                   \\
      \tableline    &             &                   &                   &                      &                         &                    &                        &                          &                              &                                   &                                   &                           &                           &                            &                         &                     &                          & \vspace{-2.5mm}                    \\
      SFHO\_145145  & 1.45/1.45   & 2.77              & 0.78              & 0.11                 & 0.06                    & 16                 & 31\,\%                 & 0.12                     & 0.14                         & 1.19                              & 4.06                              & 2.3                       & 0.23                      & 204                 & 27                      & 0/0                 & 0/0                      & 0/0                                \\
      SFHO\_1218    & 1.2/1.8     & 2.78              & 0.76              & 0.14                 & 0.06                    & 3.5                & 16\,\%                 & 0.42                     & 0.68                         & 1.60                              & 5.46                              & 2.4                       & 0.26                      & 238                 & 29                      & 0/0                 & 0/0                      & 0/0                                \\
      SFHO\_1218a3  & 1.2/1.8     & 2.78              & 0.76              & 0.14                 & 0.03                    & 3.5                & 16\,\%                 & 0.42                     & 0.68                         & 0.91                              & 3.11                              & 2.9                       & 0.12                      & 565                 & 47                      & 0/0                 & 0/0                      & 0/0                                \\
      SFHO\_1218a12 & 1.2/1.8     & 2.78              & 0.76              & 0.14                 & 0.12                    & 3.5                & 16\,\%                 & 0.42                     & 0.68                         & 2.31                              & 7.89                              & 1.9                       & 0.47                      & 93                  & 19                      & 0/0                 & 0/0                      & 0/0                                \\
      TM1\_13520    & 1.35/2.0    & 3.09              & 0.75              & 0.19                 & 0.06                    & 18                 & 5.8\,\%                & 0.21                     & 0.40                         & 2.16                              & 7.37                              & 2.3                       & 0.26                      & 283                 & 34                      & 0/0                 & 0/0                      & 0/0                                \\
      TM1\_1451     & 1.4/5.1     & 6.08              & 0.83              & 0.34                 & 0.06                    & --                 & --                     & --                       & --                           & 1.38                              & 4.71                              & 2.4                       & 0.10                      & 392                 & 72                      & 8.17/2.05           & 34/8.2                   & 0.48/2.0                           \\
      \tableline
    \end{tabular}
    \tablecomments{Each model name indicates the nuclear EOS used in the merger simulation, i.e. TM1
      \citep{Hempel2012} or SFHO \citep{Steiner2013}. Values of time-dependent quantities are given
      at the final simulation times. The sub-/superscript `pol' indicates quantities within the two
      polar cones of half-opening angle 45$^\circ$. All quantities are defined in the main text.}
\end{table*}

%
\begin{figure*}
  \includegraphics[width=\textwidth]{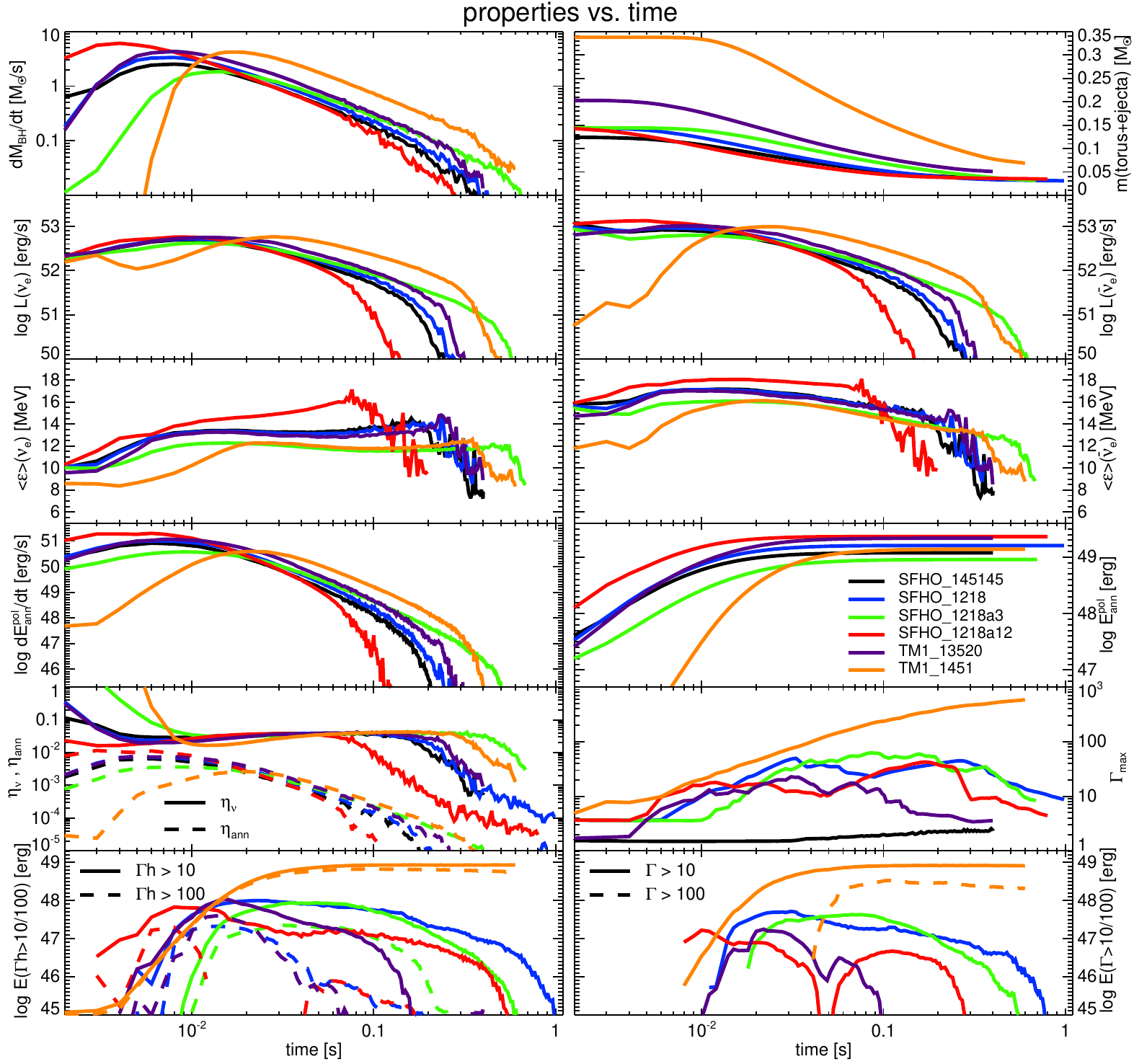}
  \caption{Time evolution of the mass-accretion rate, $\dot{M}_{\mathrm{BH}}$, torus plus ejecta
    mass, $m$, electron neutrino and antineutrino luminosities, $L_{\nu_e,\bar\nu_e}$, mean energies
    $\langle\varepsilon\rangle_{\nu_e,\bar\nu_e}$, total annihilation rate,
    $\dot{E}_{\mathrm{ann}}^{\mathrm{pol}}$, deposited annihilation energy,
    $E_{\mathrm{ann}}^{\mathrm{pol}}$, efficiencies of neutrino emission, $\eta_\nu$, and
    annihilation, $\eta_{\mathrm{ann}}$, maximum Lorentz factor, $\Gamma_{\mathrm{max}}$, and total
    energies carried by material with $\Gamma h$ or $\Gamma$ above 10 or 100,
    $E_{\Gamma h>10/100}$ and $E_{\Gamma>10/100}$, respectively. The mean energies
    are defined as $\langle\varepsilon\rangle_\nu\equiv L_\nu/L_{N,\nu}$, where $L_{N,\nu}$ is the
    total number emission rate, and $L_{\nu}, L_{N,\nu}$ are measured in the lab-frame at
    $r=500\,$km.}
  \label{fig:jet_time}
\end{figure*}

Binary neutron star (NS-NS) and neutron star-black hole (NS-BH) mergers release huge amounts of
energy in a short time and a small spatial volume. Therefore, a possible role of these events as
sources of cosmic gamma-ray bursts has been proposed already decades ago \citep{Paczynski1986,
  Eichler1989}. Well localized short gamma-ray bursts (sGRBs) provide observational support for such
a connection in particular because of their low-density environments, partially large offsets from
their galactic hosts and their occurrence also in galaxies with low star-formation activity
\citep{Nakar2007, Berger2014, Fong2015}.

NS-NS and NS-BH mergers can produce BH-torus systems as remnants, or in the former case also massive
neutron stars (MNSs), which could be transiently or long-term stable, depending on their mass
and the uncertain nuclear equation of state (EOS). The GRB is understood as consequence of highly
collimated, ultrarelativistic polar outflows or jets of low-density plasma or Poynting flux, whose
energy is partly converted to $\gamma$-rays at distances of $10^{12}$--$10^{13}$\,cm, far away from
the compact remnant \citep[e.g.][]{Piran2004}. As possible energy sources to drive these jets, the
annihilation of neutrino-antineutrino pairs radiated by the hot accretion torus is discussed
(e.g. \citealp{Eichler1989, Woosley1993, Jaroszynski1993, Ruffert1999a, Popham1999, DiMatteo2002,
  Birkl2007}, \citealp{Dessart2009}, \citealp{Zalamea2011}) or the energy release by
magnetohydrodynamic effects, Poynting flux \citep[e.g.][]{Blandford1977, McKinney2009,
  Paschalidis2015} or electron-positron pair production \citep{Usov1992} associated with ultrastrong
magnetic fields threading the BH-torus system or MNS and their environments.

Individual aspects and special questions of these scenarios have been intensively investigated
previously, for example the emergence of a jet-favorable magnetic field configuration in NS-NS/BH
mergers \citep{Paschalidis2015, Kiuchi2015, Dionysopoulou2015}, the efficiency of neutrino
production in accretion disks and tori with steady-state accretion rate \citep[e.g.][]{Popham1999,
  DiMatteo2002, Chen2007}, the efficiency of neutrino-antineutrino ($\nu\bar\nu$-) annihilation
above the poles of a BH \citep[e.g.][]{Birkl2007, Zalamea2011, Richers2015}, or the jet acceleration
and collimation by the interaction with the accretion torus, non-relativistic winds \citep{Aloy2005,
  Murguia-Berthier2014} or the ejecta from dynamical NS-NS mergers \citep{Nagakura2014,
  Duffell2015}. Also, time-dependent and self-consistent hydrodynamic simulations of BH-torus
remnants including some neutrino treatment have been performed to study neutrino emission and
annihilation \citep{Ruffert1999a, Setiawan2004, Setiawan2006, Fernandez2013, Just2015a,
  Foucart2015}.

In this work we focus on the neutrino-powered GRB scenario. For the first time, we simulate (in
axisymmetry) the evolution of BH-torus systems including (magnetic-field related) viscosity effects,
energy-dependent neutrino transport and $\nu\bar\nu$-annihilation. Moreover, we include
self-consistently the formation of relativistic polar outflows and their interaction with the
rapidly expanding, subrelativistic ejecta shell that is dynamically expelled during binary NS
merging. The considered ejecta and remnant conditions (masses, spins, expansion velocities,
energies) are set up in accordance with relativistic NS-NS and NS-BH merger simulations. The goal of
our study is to investigate the requirements for neutrino-powered GRB jets.

\begin{figure*}
  \includegraphics[width=\textwidth]{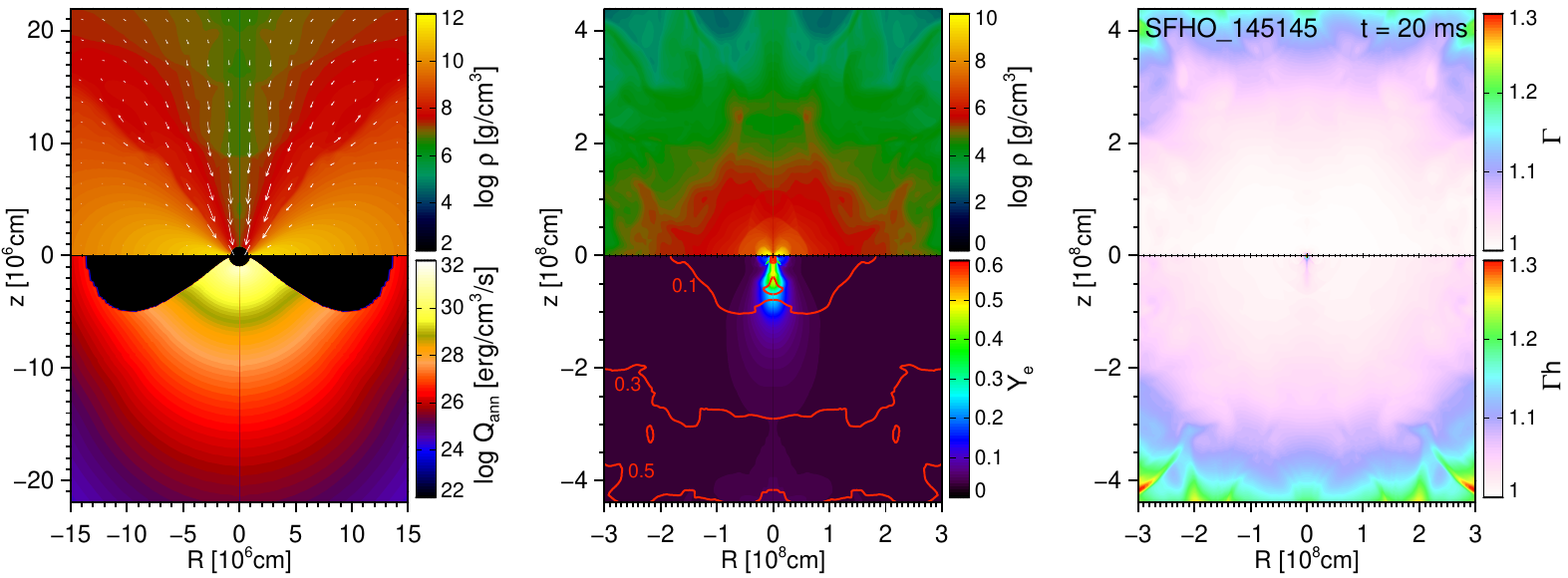}
  \includegraphics[width=\textwidth]{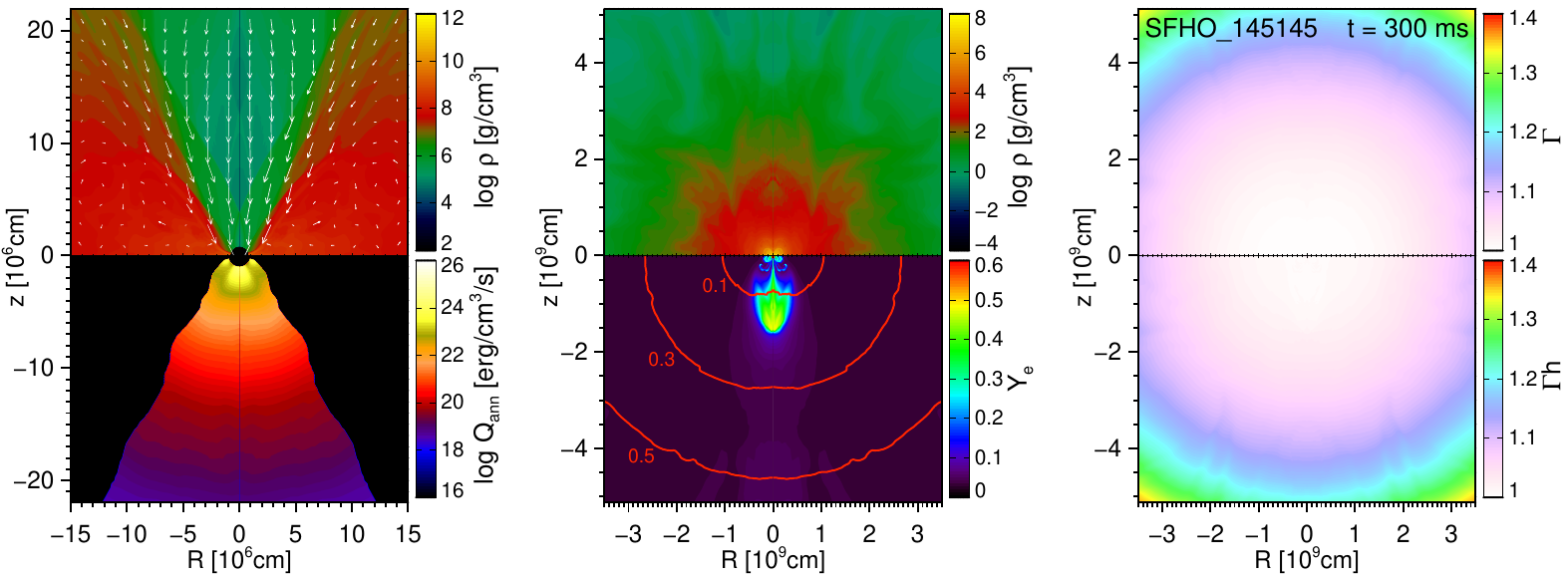}
  \caption{Maps of density, $\rho$, annihilation rate, $Q_{\mathrm{ann}}$, electron fraction, $Y_e$,
    Lorentz factor, $\Gamma$, and terminal Lorentz factor, $\Gamma h$, for NS-NS merger model
    SFHO\_145145 at the two times indicated in the right panel of each row. Note the different
    spatial and color scales for different times. Arrows indicate the poloidal (i.e. projected into
    the $R$-$z$ plane) velocity vectors. Their length is limited to the distance between two
    neighboring arrows, which corresponds to 0.2\,$c$. The red lines are isocontours of the poloidal
    velocity at values labeled in units of $c$.}
  \label{fig:cont_SFHO145145}
\end{figure*}
\begin{figure*}
  \includegraphics[width=\textwidth]{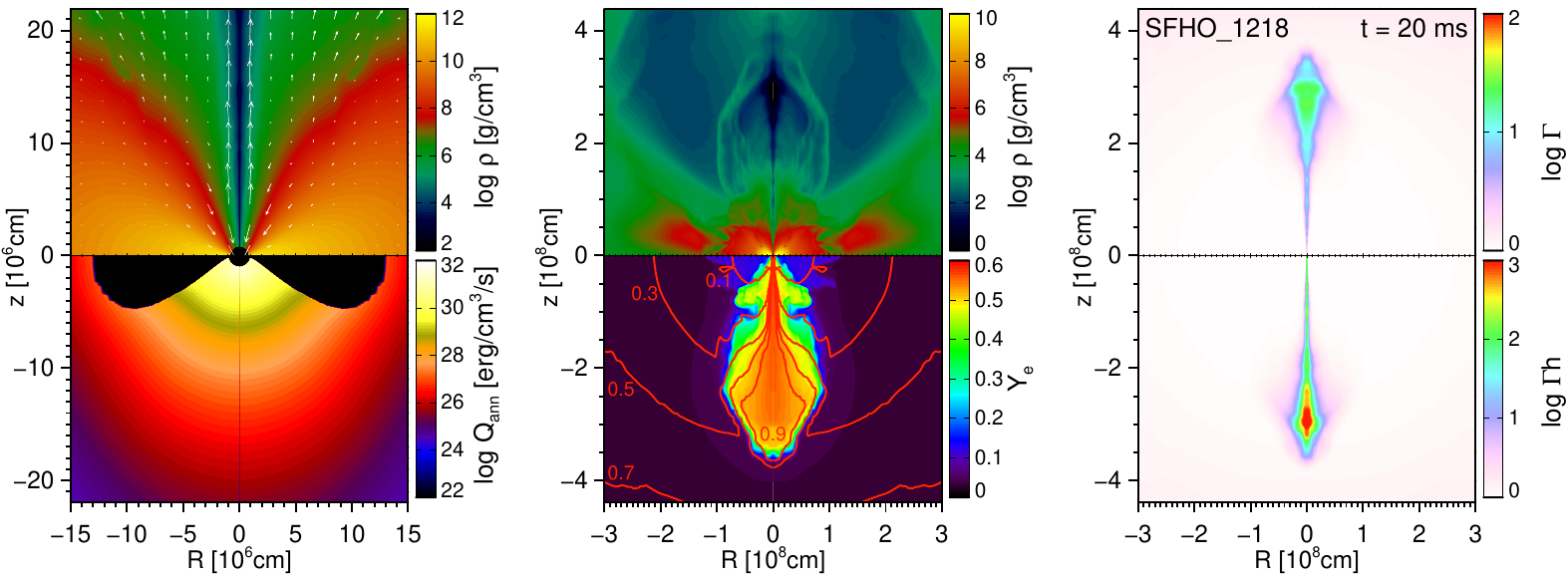}
  \includegraphics[width=\textwidth]{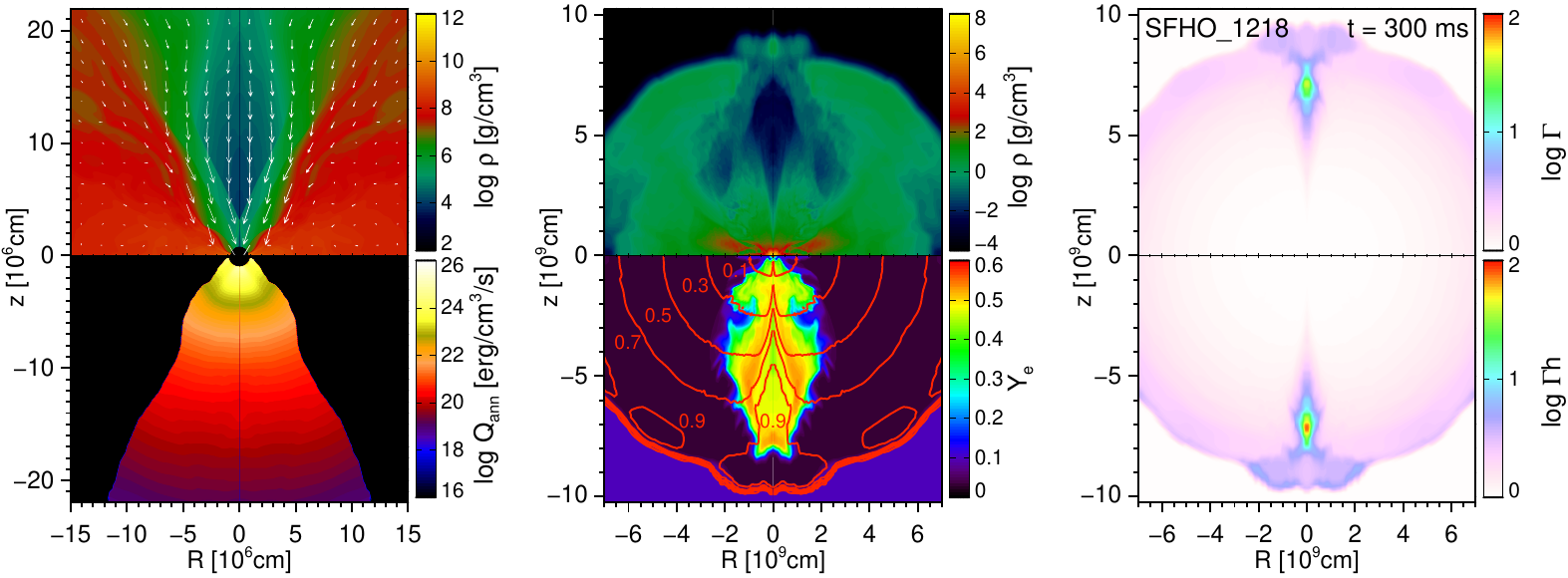}
  \caption{Same as Fig.~\ref{fig:cont_SFHO145145} but for model SFHO\_1218 and with partially
    different spatial and color scales.}
  \label{fig:cont_SFHO1218}
\end{figure*}
\begin{figure*}
  \includegraphics[width=\textwidth]{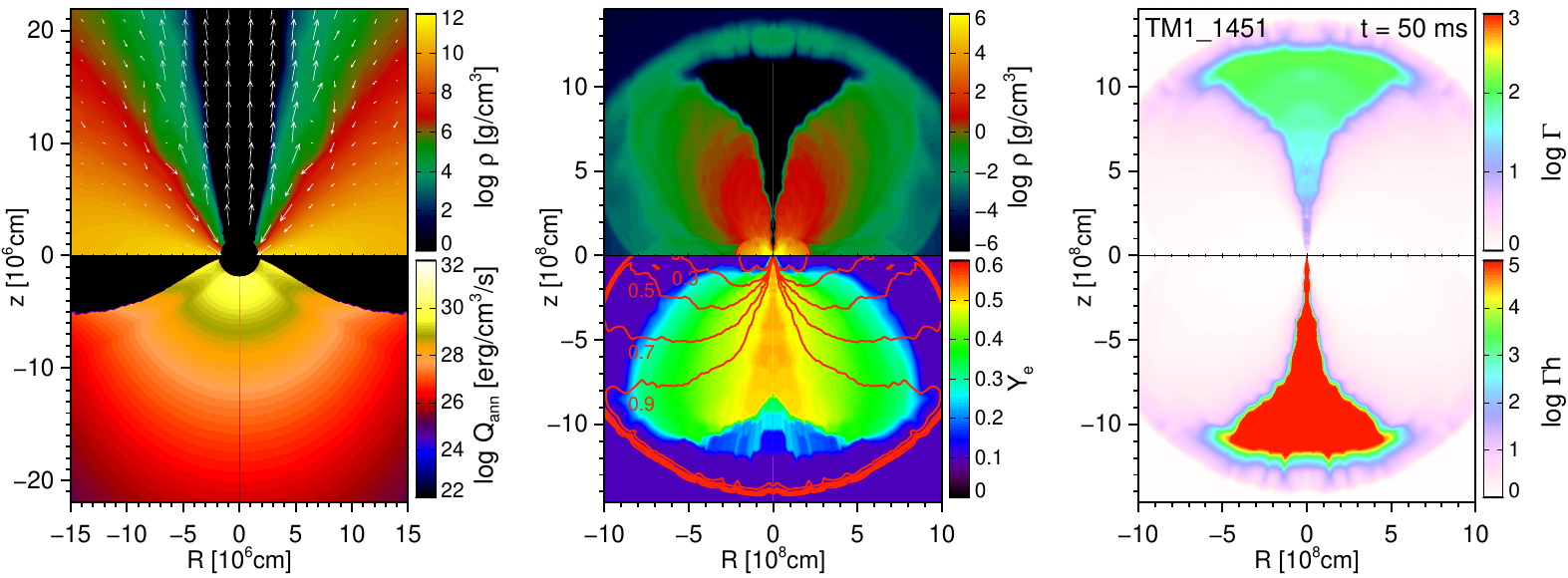}
  \includegraphics[width=\textwidth]{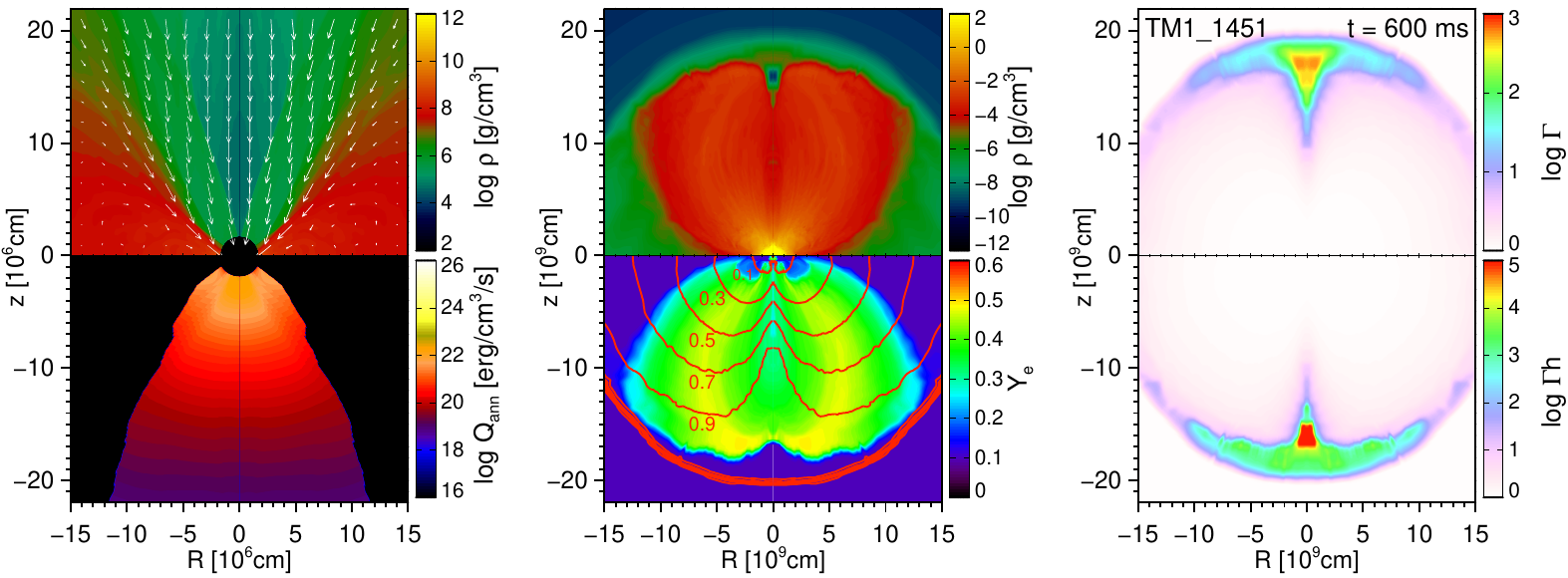}
  \caption{Same as Fig.~\ref{fig:cont_SFHO145145} but for NS-BH merger model TM1\_1451 and with
    partially different spatial and color scales.}
  \label{fig:cont_TM11451}
\end{figure*}

\section{Numerical treatment and model setup}
The simulations were performed with the ALCAR code \citep{Just2015} that solves the equations of
hydrodynamics along with a multi-energy group M1-type scheme for the neutrino transport. Most input
physics is adopted from BH-torus simulations presented in \citet{Just2015a}. However, we now use a
special relativistic instead of a Newtonian solver for the Euler equations, allowing us to
accurately follow relativistic jet outflows. Note that in the bulk of the disk the conditions are at
most mildly relativistic, hence the torus evolution and the properties of the subrelativistic ejecta
as described in \cite{Just2015} remain essentially unaffected by switching to the relativistic
solver. For the same reason we keep using the Newtonian version of the viscosity terms.

The transport of electron neutrinos and antineutrinos employs 10 energy groups distributed
logarithmically within $[0,80\,\mathrm{MeV}]$ and source terms for absorption, emission and
scattering interactions with free nucleons (as given in \citealp{Bruenn1985}). The velocities
entering the neutrino transport are subject to the restrictions explained in
\cite{Just2015a}. Energy- and momentum-transfer rates due to $\nu\bar\nu$-annihilation are
calculated from the annihilation-rate 4-vector (\citealp{Birkl2007}, typos corrected by
\citealp{Zalamea2011}) expressed in terms of the evolved neutrino moments \citep[analogously to
Eq.~10 of][]{Dessart2009} and are used as source terms for the hydrodynamic equations. Since we are
only interested in the effects of annihilation near the baryon-poor funnel, we ignore annihilation
in regions where the density $\rho > 10^{11}\,$g\,cm$^{-3}$ or where neutrino cooling dominates
heating.

We apply a microphysical equation of state (including a 4-species baryon gas, electrons, positrons
and photons), the pseudo-Newtonian gravitational potential by \citet{Artemova1996}, and a viscosity
scheme using $\alpha_{\mathrm{vis}}$ to parametrize magnetohydrodynamic, turbulent angular momentum
transport (\citealp{Shakura1973}; in the version denoted as ``type 2'' in \citealp{Just2015a}). The
initial torus equilibrium configuration is computed as described in \citet{Just2015a} for given
initial mass, $M_{\mathrm{BH}}$, and dimensionless spin parameter, $A_{\mathrm{BH}}$, of the BH, and
torus mass, $m_{\mathrm{torus}}$ (see \citealp{Just2015a} and references therein for details). For
these quantities we adopt values guided by merger simulations that were performed \citep[and
partially discussed in][]{Just2015a} with a relativistic smoothed-particle hydrodynamics (SPH) code
using the conformal flatness condition \citep{Oechslin2002, Bauswein2010b}.

To take into account the envelope of material -- mostly containing dynamical ejecta -- around the
BH-torus system formed during the merger, we map the azimuthally averaged distributions of density,
electron fraction, pressure, and velocity from the corresponding original merger model onto the grid
regions surrounding our manually constructed torus. We utilize the SPH configurations resulting at
about $\sim$1--5\,ms after the BH formation and identify the isotropic coordinate radius with our
radial coordinate $r$ \citep[see][for details concerning the dynamical
ejecta]{Bauswein2013}. However, we disregard all material from the merger model for radii $r<50\,$km
because, first, the inconsistency between both radius versions grows for smaller radii, and second,
because the original SPH torus is replaced by our self-constructed torus. The remaining volume
surrounding the torus and envelope is initially filled with a dynamically irrelevant amount of cold
gas having a density of $\rho=1.5$\,g\,cm$^{-3}$ for $r\leq 10^3\,$km and $\propto r^{-4}$ for
higher radii.

We assume equatorial and axi-symmetry and use spherical polar coordinates. However, although we
simulate only one hemisphere we always take into account both hemispheres for volume
integration. The radial grid consists of 800 zones of exponentially increasing width and extends
from $\sim$(1--2)$\times 10^6\,$cm up to $\sim$3$\times 10^{11}\,$cm. The angular grid consists of
120 zones linearly distributed within the domain $[0^\circ, 30^\circ]$ and another 120 zones of
exponentially increasing width covering $[30^\circ,90^\circ]$. We validated numerical convergence of
our main results.

We consider three NS-NS merger models (Table~\ref{tab:jetmodels}). Model SFHO\_145145 is
characterized by symmetric progenitor masses, $M_1, M_2$, and a delayed ($\sim$10\,ms) BH formation
after the merger, while models SFHO\_1218 and TM1\_13520 result from asymmetric mergers and prompt
BH formation. Hence \citep{Bauswein2013}, compared to the former model both latter models exhibit a
lower total envelope mass, $m_{\mathrm{env}}$, and a reduced fraction of mass,
$m_{\mathrm{env,pol}}/m_{\mathrm{env}}$, located in the two polar cones with $45^\circ$ half-opening
angle ($m_{\mathrm{env,pol}}/m_{\mathrm{env}}\approx 0.29$ would result for an isotropic
distribution), while the corresponding average velocities,
$\bar{v}_{\mathrm{env}}, \bar{v}_{\mathrm{env,pol}}$, are higher (Table~\ref{tab:jetmodels}). Two
models with different values of $\alpha_{\mathrm{vis}}$ are added to investigate the dependence on
viscosity. For model TM1\_1451 linked to a NS-BH merger (Table~\ref{tab:jetmodels}) we do not map
any material from the merger simulation onto our grid, because in NS-BH mergers the dynamical ejecta
are almost exclusively expelled in the equatorial direction \citep[e.g.][]{Just2015a} and therefore
have a strongly reduced impact on the jet compared to NS-NS mergers.

To assess the efficiency of $\nu\bar\nu$-annihilation for most optimistic cases, we employ favorable
rather than typical merger configurations (except for SFHO\_145145, e.g. \citealp{Dominik2012,
  Lattimer2012}), slightly overestimated torus masses \citep[e.g.][]{Kyutoku2015}, and a neutrino
treatment that tends to overestimate the $\nu\bar\nu$-annihilation rates near the poles \citep[see
the comparison of M1 with ray-tracing results in][]{Just2015a}.

\section{Results}

\subsection{Torus evolution, neutrino emission and annihilation}

The time evolution of several quantities characterizing the $\nu\bar\nu$-annihilation is illustrated
in Fig.~\ref{fig:jet_time}. All modeled tori traverse the two evolutionary phases denoted as NDAF
and ADAF (neutrino- and advection-dominated accretion flow, respectively; see e.g.,
\citealp{Metzger2008c, Fernandez2013, Just2015a}). In the NDAF phase the temperatures and thus
neutrino production rates are high enough for viscous heating to be efficiently compensated by
neutrino emission, while in the subsequent ADAF phase neutrino reactions cease and ultimately freeze
out owing to low temperatures. During the NDAF phase torus mass accreted onto the BH per unit of
time, $\dot{M}_{\mathrm{BH}}$, is converted into energy-emission rates (i.e. luminosities) of both
neutrino species, $L_{\nu_e,\bar\nu_e}$, with an efficiency
$\eta_\nu \equiv (L_{\nu_e}+L_{\bar\nu_e})/(\dot{M}_{\mathrm{BH}}c^2)$ (with the speed of light $c$)
of a few per cent (Fig.~\ref{fig:jet_time}). After being emitted, a fraction of neutrinos
pair-annihilate, giving rise to a local heating rate, $Q_{\mathrm{ann}}$
(Figs.~\ref{fig:cont_SFHO145145}--~\ref{fig:cont_TM11451}), which is highest close to the BH, since
here the neutrino densities are largest, and steeply decreases with radius. The total annihilation
rate, $\dot{E}_{\mathrm{ann}}^{\mathrm{pol}}\equiv \int_{V,45} Q_{\mathrm{ann}}\mathrm{d}V$ (with
the integral performed only in the two polar cones with $45^\circ$ half-opening angle), is roughly
proportional to $L_{\nu_e}L_{\bar\nu_e}$, multiplied by additional factors accounting for the
geometry of the emitting region and the preference for high neutrino mean energies
$\langle\varepsilon\rangle_\nu$ \citep[see, e.g.,][]{Goodman1987, Setiawan2006}. The annihilation
efficiency,
$\eta_{\mathrm{ann}} \equiv \dot{E}_{\mathrm{ann}}^{\mathrm{pol}} / (L_{\nu_e}+L_{\bar\nu_e})$, is
therefore approximately proportional to $L_{\nu_e}$ (noting that $L_{\bar\nu_e}\sim L_{\nu_e}$).
Because $L_{\nu}$ decreases roughly like $\dot{M}_{\mathrm{BH}}$ (Fig.~\ref{fig:jet_time}),
$\eta_{\mathrm{ann}}$ declines earlier than $\eta_\nu$ and, hence, the time $T^{90}_{\mathrm{ann}}$,
at which $90\,\%$ of the final $E_{\mathrm{ann}}^{\mathrm{pol}}$ has been deposited, is
significantly shorter than the duration of the NDAF phase, $T_{\mathrm{NDAF}}$, defined as the time
when $\eta_\nu$ drops below $\simeq 1\,\%$ (Table~\ref{tab:jetmodels}).

The comparison between the models is facilitated by considering (Table~\ref{tab:jetmodels}) the
global versions of the aforementioned efficiencies,
$\bar\eta_\nu\equiv E_\nu / (m_{\mathrm{torus}}c^2)$ and
$\bar\eta_{\mathrm{ann}}\equiv E_{\mathrm{ann}}^{\mathrm{pol}} / E_\nu$ with
$E_\nu\equiv \int (L_{\nu_e}+L_{\bar\nu_e})\mathrm{d}t$, by which means the total annihilation
energy
$E_{\mathrm{ann}}^{\mathrm{pol}} = \bar\eta_\nu\,\bar\eta_{\mathrm{ann}}\, m_{\mathrm{torus}}c^2$.
The emission efficiency, $\bar\eta_\nu$, is similar for all models (up to some reduction for higher
viscosity caused by enhanced neutrino trapping at very early times). The main reason for this
similarity are the comparable values of the dimensionless BH spin, $A_{\mathrm{BH}}$, for all
models, which determine the amount of specific gravitational energy that can be released by gas
before being accreted. Since approximately
$\eta_{\mathrm{ann}}\propto L_\nu \propto \dot{M}_{\mathrm{BH}}$ we infer that
$\bar\eta_{\mathrm{ann}}$ should grow for higher values of $m_{\mathrm{torus}}$ and
$\alpha_{\mathrm{vis}}$, and a lower value of $M_{\mathrm{BH}}$. This turns out to be broadly
consistent with our results. In the NS-BH model the positive effect of the high torus mass is more
than counterbalanced by the reduced neutrino mean energies, $\langle\varepsilon\rangle_\nu$,
compared to the NS-NS models (Fig.~\ref{fig:jet_time}) and by the high BH mass, which leads to a
longer accretion timescale.

Finally, the upper limit of energy available for any potential jet, represented by
$E_{\mathrm{ann}}^{\mathrm{pol}}$, is given by
$\sim (10^{-5}-10^{-4})\times m_{\mathrm{torus}}c^2 \sim (1-2)\times 10^{49}$\,erg in our models.
The corresponding isotropic equivalent energies are almost independent of the polar angle (because
of the approximate isotropy of the annihilation rates, see
Figs.~\ref{fig:cont_SFHO145145}--~\ref{fig:cont_TM11451}) with
$E_{\mathrm{ann}}^{\mathrm{iso}} \approx E_{\mathrm{ann}}^{\mathrm{pol}}/(1-\cos\,45^\circ)\sim
(3-8)\times 10^{49}$\,erg (Table~\ref{tab:jetmodels}).

\subsection{Polar outflows and ejecta interaction}

If and how much of $E_{\mathrm{ann}}^{\mathrm{pol}}$ can finally end up in an ultrarelativistic
outflow depends crucially on the distribution of matter surrounding the torus. This is illustrated
in Figs.~\ref{fig:cont_SFHO145145}--~\ref{fig:cont_TM11451} for three representative models, of
which each exhibits a qualitatively different outflow evolution.

In model SFHO\_145145 (Fig.~\ref{fig:cont_SFHO145145}) the amount of baryonic pollution near the
symmetry axis is so high that neutrino irradiation has hardly any impact, besides slightly
accelerating the already expanding dynamical ejecta and raising their electron fraction, $Y_e$. The
high densities are too prohibitive for annihilation to form a baryon-poor funnel, which is
ultimately needed to produce high Lorentz factors.
        
In models SFHO\_1218 (Fig.~\ref{fig:cont_SFHO1218}) and TM1\_13520 the configuration is better
suited for the development of relativistic outflows because the poles are loaded with relatively
small amounts of matter. Hence, in both models funnels are successfully created, allowing jets to
form and to expand outwards along the axis. Similar to jets in collapsars
\citep[e.g.][]{MacFadyen1999, Bromberg2011}, the jets consist of thin, ultrarelativistic beams
surrounded by cocoons and mildly relativistically propagating heads. Right after the jets are
launched, essentially all annihilation energy dumped into the funnels is used to power the beams,
immediately increasing $\Gamma h$ (where $\Gamma$ and $h$ are Lorentz factor and specific enthalpy,
respectively), which for adiabatic expansion represents an estimate for the terminal Lorentz
factor. Therefore, the energy carried by material with $\Gamma h > 10$ and 100,
$E_{\Gamma h>10/100}$, rises. The subsequent expansion of beam material then allows thermal energy
to be converted into kinetic energy, leading to a growth also of $E_{\Gamma>10/100}$, the energy of
material with $\Gamma>10/100$ (see Fig.~\ref{fig:jet_time} for the time evolution of
$E_{\Gamma h>10/100}$, $E_{\Gamma>10/100}$). However, the beam cannot expand freely as long as the
jet is enveloped by dynamical ejecta. Hence, while being fed with annihilation energy at its base,
the jet continually loses energy by drilling through the envelope. This happens at the expense of
beam energy, which is transported to and dissipated at the much more slowly moving head. Once
dissipated, this part of the energy is advected into the cocoon and is ultimately lost as potential
contribution to a relativistic outflow. That is, only the fraction of original annihilation energy
that is induced into the beam shortly before and after the time of an eventual break-out from the
envelope could possibly be released within a relativistic outflow. However, in the considered NS-NS
merger models the jet is not energetic enough to break out, which is why $E_{\Gamma h>10/100}$
successively decreases and finally vanishes (Figs.~\ref{fig:jet_time},~\ref{fig:cont_SFHO1218}).

Model TM1\_1451, associated with a NS-BH merger, offers the most optimistic configuration for
relativistic outflows, because the polar regions are essentially free of dynamical ejecta. However,
neutrino-driven winds and the expansion of torus matter by viscous or other effects could still
prohibit or impair the emergence of relativistic ejecta. Nevertheless, model TM1\_1451 develops a
powerful relativistic wind that contains 61\,\% (48\,\%) of the original annihilation energy in the
form of $E_{\Gamma h>10}$ ($E_{\Gamma h>100}$), as measured at $t=0.1$\,s
(Fig.~\ref{fig:jet_time}). The radius-dependent half-opening angle of the outflow, $\theta$,
reflects the temporal change of the torus configuration (Fig.~\ref{fig:cont_TM11451}): At early
times the torus is rather compact and can barely collimate the outflow, while subsequently the outer
torus layers undergo viscous and neutrino-driven expansion and form a funnel, of which the walls
collimate the outflow. Hence, the outflow consists of an ultrarelativistic ($\Gamma > 100$) core
surrounded by less relativistic ($10<\Gamma<100$) wings (see Table~\ref{tab:jetmodels} for the
corresponding energies and half-opening angles). Interestingly, the collimation leads to a
remarkable enhancement of the isotropic-equivalent energy of the ultrarelativistic core,
$E_{\Gamma>100}^{\mathrm{iso}}\approx 2\times 10^{50}\,$erg, compared to that of annihilation,
$E_{\mathrm{ann}}^{\mathrm{iso}}\approx 5\times 10^{49}\,$erg.

The distribution of $\Gamma$ and the opening angles can still change somewhat after the final
simulation time of $t=0.6\,$s because of a sizable fraction of the energy still residing in internal
energy. However, the coarser grid at higher radii renders the late evolution more prone to numerical
diffusion, which is why $E_{\Gamma>100}$ decreases at late times and we stopped the simulation.

\subsection{Conclusions}

We investigated the ability of $\nu\bar\nu$-annihilation to represent the dominant agent for
generating sGRB viable outflows. In contrast to previous studies, which either computed the
annihilation rates on a stationary background without any hydrodynamic feedback
\citep[e.g.][]{Setiawan2006, Birkl2007, Richers2015} or which launched the outflow artificially
\citep[e.g.][]{Aloy2005, Nagakura2014, Duffell2015}, we self-consistently followed the combined
neutrino-hydrodynamic evolution of the BH-torus system, the eventual launch of the jet in response
to heating by $\nu\bar\nu$-annihilation, and the jet propagation through the envelope of dynamical
ejecta produced during the binary merger. Our examined set of models covers the most relevant NS-NS
and NS-BH configurations in the sense that they are both promising regarding the annihilation
mechanism but still realistic enough to be abundantly present in nature.

We find that the total energy provided by annihilation, $E_{\mathrm{ann}}^{\mathrm{pol}}$, amounts
to a fraction of about 10$^{-5}$--10$^{-4}$ of the original torus rest-mass energy. Although this
fraction grows with torus mass and viscosity, and for lower BH mass, we find typical values of
$E_{\mathrm{ann}}^{\mathrm{pol}}$ not exceeding a few $10^{49}$\,erg.

In the NS-BH merger remnant, which is essentially free of polar dynamical ejecta, annihilation
heating is efficiently translated into a relativistic outflow, whose ultrarelativistic
($\Gamma>100$) core is dynamically collimated to a half-opening angle
$\theta_{\Gamma>100}\approx 8^\circ$ and an isotropic-equivalent energy
$E_{\Gamma>100}^{\mathrm{iso}}\approx 2\times 10^{50}\,$erg. However, compared to the median values
of observed sGRBs ($E_{\mathrm{obs}}^{\mathrm{iso}}\approx 2\times 10^{51}\,$erg and
$\theta_{\mathrm{obs}}\approx 16\pm 10^\circ$, see \citealp{Fong2015}), our results suggest that the
annihilation mechanism is energetically too inefficient to explain the majority of sGRBs, but still
could account for some low-luminosity events. This conclusion is further supported by our result
that the effective time of source activity obtained for our models is
$T^{90}_{\mathrm{ann}}\la 0.1\,$s and thus much shorter than most observed times $T^{90}$ of sGRBs
\citep{Fong2015}. A lower viscosity or higher BH spin could possibly alleviate but not solve this
issue.

Moreover, our simulations suggest that the $\nu\bar\nu$-annihilation energies in NS-NS mergers are
too low to launch jets energetically enough to pierce through the envelope of dynamical ejecta, even
in the most optimistic case of prompt BH formation and very asymmetric progenitors, which results in
less baryon-polluted polar regions. If sGRBs are connected with NS-NS mergers, our results indicate
that some other, probably magnetohydrodynamic processes \citep[e.g.][]{Paschalidis2015, Kiuchi2015,
  Dionysopoulou2015} are necessary at least to create and support a baryon-poor funnel through which
the outflow can propagate without major dissipation. Once a funnel is established, however,
$\nu\bar\nu$-annihilation could still play a non-negligible role in powering the outflow. Although
we consider only NS-NS systems here where the MNS collapses rather early, our conclusions probably
also hold for longer delay times. Then an even greater envelope mass would have to be penetrated by
the jet because of the additional neutrino- and magnetically driven outflows from the MNS
\citep{Perego2014a, Siegel2014}.

We finally point out that several simplifications entered this study, e.g. the omission of accurate
general relativistic effects, the use of viscosity to mimic magnetohydrodynamic turbulence, initial
models that were not fully consistent with the merger remnants, and the remaining approximations of
the neutrino transport.

\acknowledgements OJ is grateful to Omer Bromberg for helpful discussions. We acknowledge support by
the Max-Planck/Princeton Center for Plasma Physics (MPPC), by the DFG-funded Cluster of Excellence
EXC 153 ``Origin and Structure of the Universe'', by the European Research Council through grant
CAMAP-259276, and by the Spanish Ministerio de Ciencia e Innovaci{\'o}n through grant
AYA2013-40979-P Astrof{\'i}sica Relativista Computacional. The computations were performed at the
Max-Planck Computing and Data Facility (MPCDF).


\begin{thebibliography}{49}

\bibitem[{{Aloy} {et~al.}(2005){Aloy}, {Janka}, \& {M{\"u}ller}}]{Aloy2005}
{Aloy}, M.~A., {Janka}, H., \& {M{\"u}ller}, E. 2005, \aap, 436, 273

\bibitem[{{Artemova} {et~al.}(1996){Artemova}, {Bjoernsson}, \&
  {Novikov}}]{Artemova1996}
{Artemova}, I.~V., {Bjoernsson}, G., \& {Novikov}, I.~D. 1996, ApJ, 461, 565

\bibitem[{{Bauswein} {et~al.}(2013){Bauswein}, {Goriely}, \&
  {Janka}}]{Bauswein2013}
{Bauswein}, A., {Goriely}, S., \& {Janka}, H.-T. 2013, \apj, 773, 78

\bibitem[{Bauswein(2010)}]{Bauswein2010b}
Bauswein, A.~O. 2010, PhD Thesis, Technische Universit\"at M\"unchen,
  M\"unchen

\bibitem[{{Berger}(2014)}]{Berger2014}
{Berger}, E. 2014, \araa, 52, 43

\bibitem[{{Birkl} {et~al.}(2007){Birkl}, {Aloy}, {Janka}, \&
  {M{\"u}ller}}]{Birkl2007}
{Birkl}, R., {Aloy}, M.~A., {Janka}, H.-T., \& {M{\"u}ller}, E. 2007, \aap,
  463, 51

\bibitem[{{Blandford} \& {Znajek}(1977)}]{Blandford1977}
{Blandford}, R.~D. \& {Znajek}, R.~L. 1977, \mnras, 179, 433

\bibitem[{{Bromberg} {et~al.}(2011){Bromberg}, {Nakar}, {Piran}, \&
  {Sari}}]{Bromberg2011}
{Bromberg}, O., {Nakar}, E., {Piran}, T., \& {Sari}, R. 2011, \apj, 740, 100

\bibitem[{{Bruenn}(1985)}]{Bruenn1985}
{Bruenn}, S.~W. 1985, \apjs, 58, 771

\bibitem[{{Chen} \& {Beloborodov}(2007)}]{Chen2007}
{Chen}, W. \& {Beloborodov}, A.~M. 2007, \apj, 657, 383

\bibitem[{{Dessart} {et~al.}(2009){Dessart}, {Ott}, {Burrows}, {Rosswog}, \&
  {Livne}}]{Dessart2009}
{Dessart}, L., {Ott}, C.~D., {Burrows}, A., {Rosswog}, S., \& {Livne}, E. 2009,
  \apj, 690, 1681

\bibitem[{{Di Matteo} {et~al.}(2002){Di Matteo}, {Perna}, \&
  {Narayan}}]{DiMatteo2002}
{Di Matteo}, T., {Perna}, R., \& {Narayan}, R. 2002, \apj, 579, 706

\bibitem[{{Dionysopoulou} {et~al.}(2015){Dionysopoulou}, {Alic}, \&
  {Rezzolla}}]{Dionysopoulou2015}
{Dionysopoulou}, K., {Alic}, D., \& {Rezzolla}, L. 2015, \prd, 92, 084064

\bibitem[{{Dominik} {et~al.}(2012){Dominik}, {Belczynski}, {Fryer}, {Holz},
  {Berti}, {Bulik}, {Mandel}, \& {O'Shaughnessy}}]{Dominik2012}
{Dominik}, M., {Belczynski}, K., {Fryer}, C., {Holz}, D.~E., {Berti}, E.,
  {Bulik}, T., {Mandel}, I., \& {O'Shaughnessy}, R. 2012, \apj, 759, 52

\bibitem[{{Duffell} {et~al.}(2015){Duffell}, {Quataert}, \&
  {MacFadyen}}]{Duffell2015}
{Duffell}, P.~C., {Quataert}, E., \& {MacFadyen}, A.~I. 2015, \apj, 813, 64

\bibitem[{{Eichler} {et~al.}(1989){Eichler}, {Livio}, {Piran}, \&
  {Schramm}}]{Eichler1989}
{Eichler}, D., {Livio}, M., {Piran}, T., \& {Schramm}, D.~N. 1989, \nat, 340,
  126

\bibitem[{{Fern{\'a}ndez} \& {Metzger}(2013)}]{Fernandez2013}
{Fern{\'a}ndez}, R. \& {Metzger}, B.~D. 2013, MNRAS, 435, 502

\bibitem[{{Fong} {et~al.}(2015){Fong}, {Berger}, {Margutti}, \&
  {Zauderer}}]{Fong2015}
{Fong}, W.-F., {Berger}, E., {Margutti}, R., \& {Zauderer}, B.~A. 2015, eprint arXiv:1509.02922

\bibitem[{{Foucart} {et~al.}(2015){Foucart}, {O'Connor}, {Roberts}, {Duez},
  {Haas}, {Kidder}, {Ott}, {Pfeiffer}, {Scheel}, \& {Szilagyi}}]{Foucart2015}
{Foucart}, F., {O'Connor}, E., {Roberts}, L., {Duez}, M.~D., {Haas}, R.,
  {Kidder}, L.~E., {Ott}, C.~D., {Pfeiffer}, H.~P., {Scheel}, M.~A., \&
  {Szilagyi}, B. 2015, \prd, 91, 124021

\bibitem[{{Goodman} {et~al.}(1987){Goodman}, {Dar}, \&
  {Nussinov}}]{Goodman1987}
{Goodman}, J., {Dar}, A., \& {Nussinov}, S. 1987, \apjl, 314, L7

\bibitem[{{Hempel} {et~al.}(2012){Hempel}, {Fischer}, {Schaffner-Bielich}, \&
  {Liebend{\"o}rfer}}]{Hempel2012}
{Hempel}, M., {Fischer}, T., {Schaffner-Bielich}, J., \& {Liebend{\"o}rfer}, M.
  2012, \apj, 748, 70

\bibitem[{{Jaroszynski}(1993)}]{Jaroszynski1993}
{Jaroszynski}, M. 1993, Acta Astronomica, 43, 183

\bibitem[{{Just} {et~al.}(2015{\natexlab{a}}){Just}, {Bauswein}, {Pulpillo},
  {Goriely}, \& {Janka}}]{Just2015a}
{Just}, O., {Bauswein}, A., {Pulpillo}, R.~A., {Goriely}, S., \& {Janka}, H.-T.
  2015{\natexlab{a}}, \mnras, 448, 541

\bibitem[{{Just} {et~al.}(2015{\natexlab{b}}){Just}, {Obergaulinger}, \&
  {Janka}}]{Just2015}
{Just}, O., {Obergaulinger}, M., \& {Janka}, H.-T. 2015{\natexlab{b}}, \mnras,
  453, 3386

\bibitem[{{Kiuchi} {et~al.}(2015){Kiuchi}, {Sekiguchi}, {Kyutoku}, {Shibata},
  {Taniguchi}, \& {Wada}}]{Kiuchi2015}
{Kiuchi}, K., {Sekiguchi}, Y., {Kyutoku}, K., {Shibata}, M., {Taniguchi}, K.,
  \& {Wada}, T. 2015, \prd, 92, 064034

\bibitem[{{Kyutoku} {et~al.}(2015){Kyutoku}, {Ioka}, {Okawa}, {Shibata}, \&
  {Taniguchi}}]{Kyutoku2015}
{Kyutoku}, K., {Ioka}, K., {Okawa}, H., {Shibata}, M., \& {Taniguchi}, K. 2015,
  \prd, 92, 044028


\bibitem[{{Lattimer}(2012)}]{Lattimer2012}
{Lattimer}, J.~M. 2012, Annual Review of Nuclear and Particle Science, 62, 485

\bibitem[{{MacFadyen} \& {Woosley}(1999)}]{MacFadyen1999}
{MacFadyen}, A.~I. \& {Woosley}, S.~E. 1999, \apj, 524, 262

\bibitem[{{McKinney} \& {Blandford}(2009)}]{McKinney2009}
{McKinney}, J.~C. \& {Blandford}, R.~D. 2009, \mnras, 394, L126

\bibitem[{{Metzger} {et~al.}(2008){Metzger}, {Piro}, \&
  {Quataert}}]{Metzger2008c}
{Metzger}, B.~D., {Piro}, A.~L., \& {Quataert}, E. 2008, \mnras, 390, 781

\bibitem[{{Murguia-Berthier} {et~al.}(2014){Murguia-Berthier}, {Montes},
  {Ramirez-Ruiz}, {De Colle}, \& {Lee}}]{Murguia-Berthier2014}
{Murguia-Berthier}, A., {Montes}, G., {Ramirez-Ruiz}, E., {De Colle}, F., \&
  {Lee}, W.~H. 2014, \apjl, 788, L8

\bibitem[{{Nagakura} {et~al.}(2014){Nagakura}, {Hotokezaka}, {Sekiguchi},
  {Shibata}, \& {Ioka}}]{Nagakura2014}
{Nagakura}, H., {Hotokezaka}, K., {Sekiguchi}, Y., {Shibata}, M., \& {Ioka}, K.
  2014, \apjl, 784, L28

\bibitem[{{Nakar}(2007)}]{Nakar2007}
{Nakar}, E. 2007, \physrep, 442, 166

\bibitem[{{Oechslin} {et~al.}(2002){Oechslin}, {Rosswog}, \&
  {Thielemann}}]{Oechslin2002}
{Oechslin}, R., {Rosswog}, S., \& {Thielemann}, F.-K. 2002, Phys. Rev. D, 65,
  103005

\bibitem[{{Paczynski}(1986)}]{Paczynski1986}
{Paczynski}, B. 1986, \apjl, 308, L43

\bibitem[{{Paschalidis} {et~al.}(2015){Paschalidis}, {Ruiz}, \&
  {Shapiro}}]{Paschalidis2015}
{Paschalidis}, V., {Ruiz}, M., \& {Shapiro}, S.~L. 2015, \apjl, 806, L14

\bibitem[{{Perego} {et~al.}(2014){Perego}, {Rosswog}, {Cabez{\'o}n},
  {Korobkin}, {K{\"a}ppeli}, {Arcones}, \& {Liebend{\"o}rfer}}]{Perego2014a}
{Perego}, A., {Rosswog}, S., {Cabez{\'o}n}, R.~M., {Korobkin}, O.,
  {K{\"a}ppeli}, R., {Arcones}, A., \& {Liebend{\"o}rfer}, M. 2014, \mnras,
  443, 3134

\bibitem[{{Piran}(2004)}]{Piran2004}
{Piran}, T. 2004, Reviews of Modern Physics, 76, 1143

\bibitem[{{Popham} {et~al.}(1999){Popham}, {Woosley}, \& {Fryer}}]{Popham1999}
{Popham}, R., {Woosley}, S.~E., \& {Fryer}, C. 1999, \apj, 518, 356

\bibitem[{{Richers} {et~al.}(2015){Richers}, {Kasen}, {O'Connor},
  {Fern{\'a}ndez}, \& {Ott}}]{Richers2015}
{Richers}, S., {Kasen}, D., {O'Connor}, E., {Fern{\'a}ndez}, R., \& {Ott},
  C.~D. 2015, \apj, 813, 38

\bibitem[{{Ruffert} \& {Janka}(1999)}]{Ruffert1999a}
{Ruffert}, M. \& {Janka}, H.-T. 1999, \aap, 344, 573

\bibitem[{{Setiawan} {et~al.}(2004){Setiawan}, {Ruffert}, \&
  {Janka}}]{Setiawan2004}
{Setiawan}, S., {Ruffert}, M., \& {Janka}, H.-T. 2004, \mnras, 352, 753

\bibitem[{{Setiawan} {et~al.}(2006){Setiawan}, {Ruffert}, \&
  {Janka}}]{Setiawan2006}
---. 2006, \aap, 458, 553

\bibitem[{{Shakura} \& {Sunyaev}(1973)}]{Shakura1973}
{Shakura}, N.~I. \& {Sunyaev}, R.~A. 1973, A\&A, 24, 337

\bibitem[{{Siegel} {et~al.}(2014){Siegel}, {Ciolfi}, \&
  {Rezzolla}}]{Siegel2014}
{Siegel}, D.~M., {Ciolfi}, R., \& {Rezzolla}, L. 2014, \apjl, 785, L6

\bibitem[{{Steiner} {et~al.}(2013){Steiner}, {Hempel}, \&
  {Fischer}}]{Steiner2013}
{Steiner}, A.~W., {Hempel}, M., \& {Fischer}, T. 2013, \apj, 774, 17

\bibitem[{{Usov}(1992)}]{Usov1992}
{Usov}, V.~V. 1992, \nat, 357, 472

\bibitem[{{Woosley}(1993)}]{Woosley1993}
{Woosley}, S.~E. 1993, \apj, 405, 273

\bibitem[{{Zalamea} \& {Beloborodov}(2011)}]{Zalamea2011}
{Zalamea}, I. \& {Beloborodov}, A.~M. 2011, \mnras, 410, 2302

\end{thebibliography}

\end{document}